\documentclass[twocolumn,tighten]{aastex63}

\usepackage{amsmath}

\newcommand\latex{La\TeX}

\submitjournal{ApJ}

\shorttitle{Magnetic fluctuations and regulation of turbulence}
\shortauthors{Espinoza et al.}

\begin{document}

\title{Spontaneous Magnetic Fluctuations and Collisionless Regulation of Turbulence \\
in the Earth's Magnetotail}

\correspondingauthor{Crist{\'o}bal M. Espinoza}
\email{cristobal.espinoza.r@usach.cl}

\author[0000-0003-2481-2348]{C. M. Espinoza}
\affiliation{Departamento de F\'{\i}sica, Universidad de Santiago de Chile (USACH)}
\affiliation{Center for Interdisciplinary Research in Astrophysics and Space Exploration (CIRAS), Universidad de Santiago de Chile} 

\author[0000-0002-9161-0888]{P. S. Moya}
\affiliation{Departamento de F\'{\i}sica, Facultad de Ciencias, Universidad de Chile}


\author[0000-0002-1053-3375]{M. Stepanova}
\affiliation{Departamento de F\'{\i}sica, Universidad de Santiago de Chile (USACH)}
\affiliation{Center for Interdisciplinary Research in Astrophysics and Space Exploration (CIRAS), Universidad de Santiago de Chile} 

\author[0000-0003-3381-9904]{J. A. Valdivia}
\affiliation{Departamento de F\'{\i}sica, Facultad de Ciencias, Universidad de Chile}
\affiliation{Centro para el Desarrollo de la Nanociencia y   Nanotecnolog\'{\i}a, CEDENNA, Chile}


\author[0000-0003-0782-1904]{R. E. Navarro}
\affiliation{Departamento de F\'{\i}sica, Facultad de Ciencias F\'{\i}sicas y Matem{\'a}ticas, Universidad de Concepci{\'o}n, Chile}





\begin{abstract}
Among the fundamental and most challenging problems of laboratory, space, and astrophysical plasma physics is to understand the relaxation processes of nearly collisionless plasmas toward quasi-stationary states; and the resultant states of electromagnetic plasma turbulence. Recently, it has been argued that solar wind plasma $\beta$ and temperature anisotropy observations may be regulated by kinetic instabilities such as the ion-cyclotron, mirror, electron-cyclotron, and firehose instabilities; and that magnetic fluctuation observations are consistent with the predictions of the Fluctuation-Dissipation theorem, even far below the kinetic instability thresholds. Here, using in-situ magnetic field and plasma measurements by the THEMIS satellite mission, we show that such regulation seems to occur also in the Earth's magnetotail plasma sheet at the ion and electron scales. 
Regardless of the clear differences between the solar wind and the magnetotail environments, our results indicate that spontaneous fluctuations and their collisionless regulation are fundamental features of space and astrophysical plasmas, thereby suggesting the processes is universal.
\end{abstract}



\section{Introduction} 
\label{sec:intro}
The solar wind, the Earth's magnetosphere, and the ionosphere are an interconnected dynamical system that constitutes a great laboratory to study fundamental space plasma processes.
In its various regions, the system offers a broad variety of physical features, such as different MHD regimes, turbulent and laminar flows, energetic particle beams, and pressure gradients to name a few.
Despite remarkable efforts and valuable scientific advances, the phenomenology of the system is not fully understood and numerous problems remain unsolved \citep[e.g.][]{dbsv16,bdv+20}.

Turbulence in the magnetosphere might appear naturally as the incoming plasma flow of the solar wind encounters and interacts with the geomagnetic field.
However, the process is extremely complex and there are several unexplained phenomena that differentiate this system from an ordinary fluid wake.
Among them, the magnetotail contains regions with rather different properties, like a very turbulent plasma sheet, and two tail lobes that are less-turbulent, less dense, and less magnetic \citep[e.g.][]{akc+93,beft97,vvb+04,sa11,as+21}.

One poorly understood mechanism is the driver for turbulence in the magnetotail. 
Turbulent behavior in the plasma sheet is present even during quiet geomagnetic conditions, 
thus variations of the solar wind properties play only a mild role. 
Among the possible magnetospheric sources of turbulence there are plasma pressure gradients, particle beams, anisotropic particle distributions, magnetic reconnection, and the specific boundary conditions of the plasma sheet. 
Turbulence may be important for the generation and stability of the plasma sheet, as turbulent transport via eddy diffusion might effectively counteract the effects of plasma transport caused by the regular dawn-dusk electric field \citep{ao99,oay00,sap+09,sa11}.

Magnetospheric turbulence has been identified by studying the variability of the bulk velocity of particles \citep[e.g.][]{beft97,spva11} and also by observing magnetic fluctuations \citep[e.g.][]{vbn+04,vvb+04,wkk+05}.
On the other hand, studies of the solar wind (and also of some regions in the magnetosphere) have shown that physics at the anisotropic kinetic level can regulate plasma turbulence and the production of electromagnetic variations at the dissipation range \citep{hpt+06,bkh+09,imk13,nmm+2014,vtg+16,avmw16,yab+16,mcg+18}.

Indeed, the solar wind plasma turbulence at $1$\,AU relaxes to a dynamical quasi-stationary state in which remains for long times.  
Under this condition, in the anisotropic kinetic regime the velocity distribution of particles  can be described --as a first approximation-- by a two component distribution that is characterized by parallel ($T_\parallel$) and perpendicular ($T_\perp$) temperatures with respect to a background magnetic field $B_0$. 
The behavior of the plasma and occurrence of these parameters can be visualized in a $\beta_\parallel$-$A$ diagram; where $\beta_\parallel$ is the ratio between the parallel energy density of the particles and the magnetic energy density; and $A=T_\perp/T_\parallel$ is the thermal anisotropy \citep[e.g.][and many others]{kls+06,hpt+06,bkh+09,mkb11}. 
Measurements of the above parameters in the solar wind populate a restricted region of the diagram, an effect that can be attributed to the regulation exerted by kinetic ion-cyclotron, mirror, whistler or electron-cyclotron, and firehose instabilities in the absence of collisions and heat flux.
Thus kinetic physics could play an important role controlling the global turbulent state of  these plasmas.

Furthermore, for plasma parameters under which the system should be linearly stable, strong electromagnetic fluctuations are observed well below the instability thresholds.
This is ubiquitous for $A=1$. 
There is a number of proposed explanations for the existence of these electromagnetic fluctuations. 
One of them is that a relevant component of these fluctuations is produced by the random motion of particles in the plasma.
The process would be balanced by dissipation; so that a full understanding requires a kinetic treatment which relies on an extension of the Fluctuation-Dissipation theorem for anisotropic plasmas \citep{aam+12,nmm+2014,nmm+15,vmn+15}.

In this work we show for the first time, for both ions and electrons, that a similar kinetic regulation of turbulence and generation of magnetic fluctuations occurs also in the magnetotail plasma sheet, despite being a rather different plasma environment from the solar wind.
Indeed, while the plasma sheet is a confined plasma, the solar wind exhibits high bulk velocities.
Additionally, ion and electron energy distribution functions in the plasma sheet exhibit power law tails that are well described by $\kappa$-distributions, as opposed to the bi-Maxwellian distributions observed in the solar wind, and imply different values of the ion-to-electron temperature ratio, $T_i/T_e$ \citep{esm+18,esem+21}. 
The article is organized as follows. 
In Sections \ref{sec:data} and \ref{sec:method} we describe the data sets and methodology for obtaining the ion and electron plasma parameters, and the magnetic field fluctuation characteristics. 
In Section \ref{sec:theory} we present a brief summary of the theory of spontaneous electromagnetic fluctuations based on the fluctuation-dissipation theorem, and in Section \ref{sec:results} 
present the results of the analyses and compare theory with observations. Finally, in Section \ref{sec:conclusions} we summarize and conclude our findings.

\section{Data}
\label{sec:data}
We use in-situ measurements performed by the five satellites of the THEMIS mission \citep[Time History of Events and Macroscale Interactions during Substorms;][]{ang08} during the years 2008 and 2009. 
Data were selected from time windows that correspond to the satellite passages through the plasma sheet selected by \citet{esm+18}, and we combine the three geomagnetic conditions they considered (quiet, expansion, and recovery). 
Following their definitions, the volume of space considered was defined by the GSM coordinates $X<0$ , $|Y |<|X|$, and $|Z|<8$\,R$_E$. 
We apply no restriction to the $\beta$ parameter but, to ensure that the satellites were not in the tail lobes, the ion density was restricted to $n_i > 0.1$ cm$^{-3}$ and the ion temperature to $T_i > 1$~keV. 
To evaluate these conditions we use values averaged over 6-min-long intervals (which corresponds to half the time used by \citet{esm+18}, where the intervals overlapped by 6-min).

After having identified tens of thousands of 6-min-long intervals that satisfy the above conditions (plasma sheet crossings), each interval was divided into 120 shorter intervals of $3$\,s of duration. 
For each of these intervals we downloaded {\sl Level 2} THEMIS data consisting of high telemetry magnetic field measurements, taken by the Fluxgate Magnetometer \cite[FGM;][]{agm+08}, and {\sl BURST} mode ion and electron density and temperature measurements taken by the electrostatic analyzer \citep[ESA;][]{mcl+08}.
Three temperature components were acquired for each particle species, that correspond to one parallel and two perpendicular (to the mean magnetic field), as offered in the THEMIS ftp sites. 
While the ESA data has a time resolution of $3$\,s, FGM gives one magnetic field measurement every $\sim 0.0078$\,s.

Therefore, for each 3-s interval there are up to $N\leq384$ magnetic field measurements (each consisting of three components), one estimate of the density, and one three-component  temperature, for each species. 
We note that FGM's high telemetry data is not homogeneous and there are some gaps with no data. 
Here we report calculations only for those 3-s intervals that had $N\geq60$ FGM measurements available.

\section{Magnetic field fluctuation measurements}
\label{sec:method}
The satellite data are used to calculate magnetic field fluctuations in the direction parallel and perpendicular to the mean magnetic field direction. 
The magnetic field measurements in each 3-s interval consist of $60\leq N\leq384$ vectors $\mathbf{B}$ whose components are one parallel and two perpendicular to the average field. 
In order to account for relatively slow variations of the magnetic field --typically observed in the turbulent plasma sheet, for every 3-s interval we perform linear fits (one to each $\mathbf{B}$ component) and generate a model $\langle\mathbf{B}\rangle$.
This slowly linearly varying magnetic field is used to de-trend the individual measurements and define the more rapidly varying $d\mathbf{B}=\mathbf{B}-\langle\mathbf{B}\rangle$.

We then compute the parallel and perpendicular fluctuation levels with respect to $\langle\mathbf{B}\rangle$, as
\begin{eqnarray}
\label{eq:sigpal} {\sigma^2}_\parallel & = \frac{1}{N}\sum_i^N d\vec{\mathbf{B}}^i_\parallel \cdot d\vec{\mathbf{B}}^i_\parallel \quad, \\
\label{eq:sigper} {\sigma^2}_\perp & = \frac{1}{N}\sum_i^N d\vec{\mathbf{B}}^i_\perp \cdot d\vec{\mathbf{B}}^i_\perp \quad,
\end{eqnarray}
and define the total fluctuation level as $\sigma$, with $\sigma^2={\sigma^2}_\parallel+{\sigma^2}_\perp$.

Additionally, for each 3-s interval we calculate the average temperature in the perpendicular direction by averaging the two perpendicular values downloaded from the THEMIS sites.  
Using this in combination with the density and parallel temperature, for each 3-s interval (and for each species separately) we calculate the anisotropy parameter $A=T_\perp/T_\parallel$ and $\beta_\parallel=8\pi n k_B T_\parallel/B_0^2$, where $B_0$ is the average magnetic field amplitude during the the 3-s interval.

\section{Theory}
\label{sec:theory}
For simplicity, we assume transverse electromagnetic fluctuations propagating along the background
magnetic field $\vec{B}_0$, which seems to be a reasonable approximation in a relevant part of the $\beta$-A diagram ~\citep[see for example][]{mn2021}, in a Kappa-distributed ion-electron warm plasma with a velocity distribution function given by~\citep{Vinas2017,m+2020}
\begin{align}
\nonumber
f^{(\alpha)}(v_\perp,v_\parallel) &=\frac{n_\alpha}{\pi^{3/2}\, u_{\perp\alpha}^2\, u_{\parallel\alpha} } \frac{\Gamma(\kappa_\alpha)}{\kappa_\alpha^{1/2} \,\Gamma(\kappa_\alpha-1/2)}\\ 
&\cdot\left[1+ \frac{1}{\kappa_\alpha}\left(\frac{v_{\perp}^2}{u_{\perp\alpha}^2} + \frac{v_{\parallel}^2}{u_{\parallel\alpha}^2}\right)\right]^{-(\kappa_\alpha+1)}\,,
\label{eq:kappa}
\end{align}
where $u_{\perp \alpha} = \sqrt{2 k_B T_{\perp \alpha} / m_{\alpha}}$, $u_{\parallel \alpha} = \sqrt{2 k_B T_{\parallel \alpha} / m_{\alpha}}$, are the thermal speeds of the species $\alpha$, perpendicular and parallel to the magnetic field, respectively. Here $T_{\perp \alpha}$,  $T_{\parallel \alpha}$, $m_{\alpha}$, and $n_{\alpha}$ are their perpendicular and parallel temperatures, mass, and number density, respectively; and $k_B$ is the Boltzmann constant. Under this description, the temperature anisotropy of the species $\alpha$ is $A_\alpha = T_{\perp\alpha}/T_{\parallel\alpha}$.

The Fourier-transformed Maxwell's equations relate the circularly polarized ($+$ for right, $-$ for left) electric fields and currents through
\begin{equation}
\left(1-c^2k^2/\omega^2\right)E^{\pm}=4\pi J^{\pm}/i\omega.
\label{eq4}
\end{equation}
Although a number of hypothesis have been put forward to explain the existence of these fluctuations, it is not difficult to show that under these conditions charged particles move in random trajectories that are consistent with anisotropic velocity distributions, so that current densities and fields are also related as~\citep{nam+14,nmm+15,vmn+15}
\begin{equation}
4\pi J^\pm/i\omega=-\sum_\alpha\lambda_\alpha^{\pm}\left(E^\pm+\delta E^\pm \right).
\label{eq5}
\end{equation}
Let us note that without fluctuations, namely, ($\delta E^\pm=0$), we obtain the regular linear dispersion relation that provides frequency as a function of wavenumber $\omega(k)$. Namely,
\begin{align}
  \chi_{\alpha}^{\pm}
  &= \frac{\omega_{p \alpha}^2}{\omega^2} [R_{\alpha} +
    (\xi_{\alpha}^0 + R_{\alpha} \xi_{\alpha}^{\pm}) Z_{\kappa_{\alpha}}
    (\xi_{\alpha}^{\pm})]\,,
  \label{eq:suscep}\\
  \Lambda^{\pm} &= 1 - \frac{c^2 k^2}{\omega^2} + \sum_{\alpha}
                  \chi_{\alpha}^{\pm}\,,
                  \label{eq:disprel}
\end{align}
where $c$ is the speed of light; $\omega$ and $k$ are the frequency and wavenumber of the fluctuating fields; $\Lambda^{\pm}$ is the dispersion tensor with the sign representing the helicity of the wave; $\chi_{\alpha}^{\pm}$ is the susceptibility of ions ($\alpha=i$) and electrons ($\alpha=e$); 
$\omega_{p \alpha} = \sqrt{4 \pi n_{\alpha} q_{\alpha}^2 / m_{\alpha}}$ is the plasma frequency of the species $\alpha$, with $q_{\alpha}$ their charge; $R_{\alpha} = A_{\alpha} - 1$; 
$\xi_{\alpha}^{\sigma} = (\omega + \sigma \Omega_{\alpha}) / u_{\parallel \alpha} k$ with
$\sigma = \{ 0, \pm \}$, and $\Omega_{\alpha} = q_{\alpha} B_0 / m_{\alpha} c$ is the gyrofrequency of the species $\alpha$, respectively; and $Z_\kappa$ is the modified plasma dispersion function for Kappa-like distributions \citep{st91,hm02,nmm+15,vmn+15}. The more involved oblique treatment is left for a future study.

For anisotropic temperatures ($A\neq1$) the plasma can become unstable to the cyclotron $(A>1)$ or parallel firehose ($A+2/\beta_\parallel < 1$) instabilities~\citep{nmm+15,vmn+15}. 
The maximum growth rate of each instability can be calculated numerically through the dispersion relation $\Lambda^\pm=0$ from Eq.~\eqref{eq:disprel}. 
For its calculation we assume that most ions are protons, hence $n_i= n_p$; and treat electrons as particles with a proton-to-electron mass ratio $m_p/m_e=1836$.
The level curves of the maximum growth-rates for the different kinetic instabilities in the $\beta_\parallel$-$A$ diagram, for electrons and ions, can be fitted by an analytic function:
\begin{equation}
  \label{eq:fit}
  A_\alpha = a + \frac{b}{(\beta_{\alpha\parallel} - \beta_0)^c}\,. 
\end{equation}
This equation is slightly generalized from \cite{hpt+06} with $a\neq1$, since instabilities with
$\gamma_{\rm{max}}/\Omega_\alpha>0.01$ do not necessarily converge to
$A_\alpha=1$ when $\beta_{\alpha\parallel}\gg1$. 
The fitted parameters $a,\,b,\,c$, and $\beta_0$ of Eq.~\eqref{eq:fit}, for some given contours $\gamma_{\rm{max}}/\Omega_\alpha$ in the $\beta_\parallel$-$A$ diagram for ions and electrons, are listed in Tables \ref{tab:fiti} and \ref{tab:fite}, respectively. 

\begin{deluxetable}{lcccc}
\tablecaption{Fitted ion parameters for Eq. \eqref{eq:fit} \label{tab:fiti}}
\tablehead{
	\colhead{$\gamma_{\rm{max}}/\Omega_i$} & \colhead{$a$} & \colhead{$b$} 
		& \colhead{$\beta_0$} & \colhead{$c$} }
\startdata
    \multicolumn{5}{c}{Ion-cyclotron (Alfv\'en) instability}\\ 
    $10^{-4}$ & $0.9877$ & $0.2775$ & $-0.0013$ & $0.4260$  \\
    $10^{-3}$ & $0.9958$ & $0.3636$ & $-0.0011$ & $0.4231$  \\
    $10^{-2}$ & $1.0421$ & $0.5250$ & $-0.0006$ & $0.4301$  \\
    $10^{-1}$ & $1.3655$ & $1.0190$ & $0.0013$ & $0.4956$   \\ \hline
    \multicolumn{5}{c}{Ion-Firehose instability} \\ 
    $10^{-4}$ & $1.0360$ & $-0.3214$ & $0.2079$ & $0.4322$ \\
    $10^{-3}$ & $1.0498$ & $-0.4229$ & $0.3213$ & $0.4353$ \\
    $10^{-2}$ & $1.0344$ & $-0.5791$ & $0.4351$ & $0.5292$ \\
    $10^{-1}$ & $0.8559$ & $-1.1457$ & $0.1567$ & $0.9505$ \\
\enddata
\tablecomments{Values are presented for different maximum
    growth rates $\gamma_{\rm{max}}/\Omega_i$ (normalized to the ion
    gyrofrequency) for the ion-cyclotron and ion-firehose
    instabilities. 
    Valid for $0.01 < \beta_{i\parallel} < 100.0$,
    $0.1 < A_i < 10.0$, with $\beta_{e\parallel}=\beta_{i\parallel}/5$,
    $\kappa_i=7$, $\kappa_e=5$, $A_e=1$, $m_p/m_e=1836.0$, and $\omega_{pe}/\Omega_e=20$.
    }
\end{deluxetable}

\begin{deluxetable}{lcccc}
\tablecaption{Fitted electron parameters for Eq. \eqref{eq:fit} \label{tab:fite}}
\tablehead{
	\colhead{$\gamma_{\rm{max}}/\Omega_e$} & \colhead{$a$} & \colhead{$b$} 
		& \colhead{$\beta_0$} & \colhead{$c$} }
\startdata
    \multicolumn{5}{c}{Electron-cyclotron (whistler) instability}\\ 
    $10^{-4}$ & $0.9949$ & $0.0897$ & $-0.0043$ & $0.6320$ \\
    $10^{-3}$ & $0.9912$ & $0.1573$ & $-0.0038$ & $0.5793$ \\
    $10^{-2}$ & $0.9938$ & $0.3094$ & $-0.0022$ & $0.5234$ \\
    $10^{-1}$ & $1.0661$ & $0.8156$ & $0.0012$ & $0.5113$  \\ \hline
    \multicolumn{5}{c}{Firehose instability} \\ 
    $10^{-4}$ & $0.9465$ & $-1.3415$ & $0.0229$ & $0.9780$ \\
\enddata
\tablecomments{Values are presented for different maximum
    growth rates $\gamma_{\rm{max}}/\Omega_e$ (normalized to the electron
    gyrofrequency) for the electron-cyclotron and electron-firehose
    instabilities. 
    Valid for
    $0.01 < \beta_{e\parallel} < 100.0$, $0.1 < A_e < 10.0$, and
    $A_i=1$.  
    Other parameters are the same as in Table~\ref{tab:fiti}.
    }
\end{deluxetable}

The presence of random fluctuations, however, breaks the linear restriction for $\omega$ and $k$, producing a continuum spectrum, and we need an additional relation to close the system of equations. To do so, we appeal to statistical mechanics and the Fluctuation-Dissipation Theorem~\citep{nam+14,nmm+15,vmn+15}. With this approach we can estimate the ensemble averaged fluctuating (perturbed) fields from
\begin{equation}
    \langle \mathbf{E} \rangle =\frac{\sum_{\alpha} \int dx f^{(\alpha)} \mathbf{E}}{\sum_\alpha \int dx f^{(\alpha)}},
\end{equation}
where $f^{(\alpha)}$ is the particle distribution function for species $\alpha$ that depends on the parallel and perpendicular momentum components. In the presence of transverse fluctuations we perturb the perpendicular component of the momentum as $\mathbf{p}_\perp\to\mathbf{p}_{\perp}-q\delta \mathbf{A}_\perp/c$, with $\delta \mathbf{A}_\perp=c\delta \mathbf{E}_\perp/i\omega$, in the above expression. Expanding for small $\delta \mathbf{A}_\perp$, and assuming that $\langle E \rangle_0=0$ over the unperturbed distribution functions, we arrive at
\begin{equation}
  \langle \mathbf{E} \rangle  = \frac{\sum_\alpha \int dx (\nabla_{\mathbf{p}_\perp} f^{(\alpha)})\cdot( q_\alpha \delta \mathbf{A}^*_\perp/c) \mathbf{E}}{\sum_\alpha \int dx f^{(\alpha)}},
  \label{eq:espec}
\end{equation}
that produces terms proportional to $\sum_\ell\langle \mathbf{E} J_{\ell}\rangle\delta A^*_\ell$, where the $\ell$ summation is over the perpendicular components (see~\citep{nmm+15} for more details). Using Eqs.~(\ref{eq4})-(\ref{eq:disprel}), it is possible to find an expression for  $\langle | E_{k\omega}^{\pm} |^2 \rangle$. In the case of $\kappa$-distributions with $f^{(\alpha)}$ given by Eq.~\eqref{eq:kappa}, the spectrum of magnetic fluctuations, after transforming the electric field fluctuations into magnetic fluctuations with the help of Faraday's law, is
\begin{equation}
        \frac{ \langle | B_{k\omega}^{\pm} |^2 \rangle }{8 \pi} = \frac{c^2 k^2/\omega}{\omega^2 - c^2 k^2} 
  	\sum_{\alpha} \frac{\kappa_{\alpha} k_B T_{\perp \alpha} }{\kappa_{\alpha} - 3 / 2} 
    {\rm Im} \left\{ \frac{\chi_{\alpha}^{\pm}}{\Lambda^{\pm}} \right\}\,,
\end{equation}
where the summation considers ions and electrons ($\alpha=e,i$). Finally, to calculate an estimate of the total magnetic power, we evaluate the normalized quantity
\begin{equation}
  \label{eq:WB}
  W_B = \frac{n_p v_A}{\Omega_p} \int d\omega \int dk\, \frac{\langle | B_{k\omega}^{\pm} |^2 \rangle }{B_0^2}\,,
\end{equation}
where $n_p$ and $\Omega_p$ are the density and gyro-frequency of protons, and $v_A=B_0/\sqrt{4\pi n_p m_p}$ is the Alfv\'en speed. In the following sections, Eq.~\eqref{eq:WB} will be evaluated for different values of $\beta_{\alpha\parallel}$ and $A_\alpha$ for both protons ($\alpha=i$) and electrons ($\alpha=e$), and compared with satellite data of the observed perpendicular magnetic fluctuations through Eq.~\eqref{eq:sigper}.

\section{Results}
\label{sec:results}

\begin{figure*}
	\includegraphics[trim=0 0 105 10, clip, height=7.1cm]{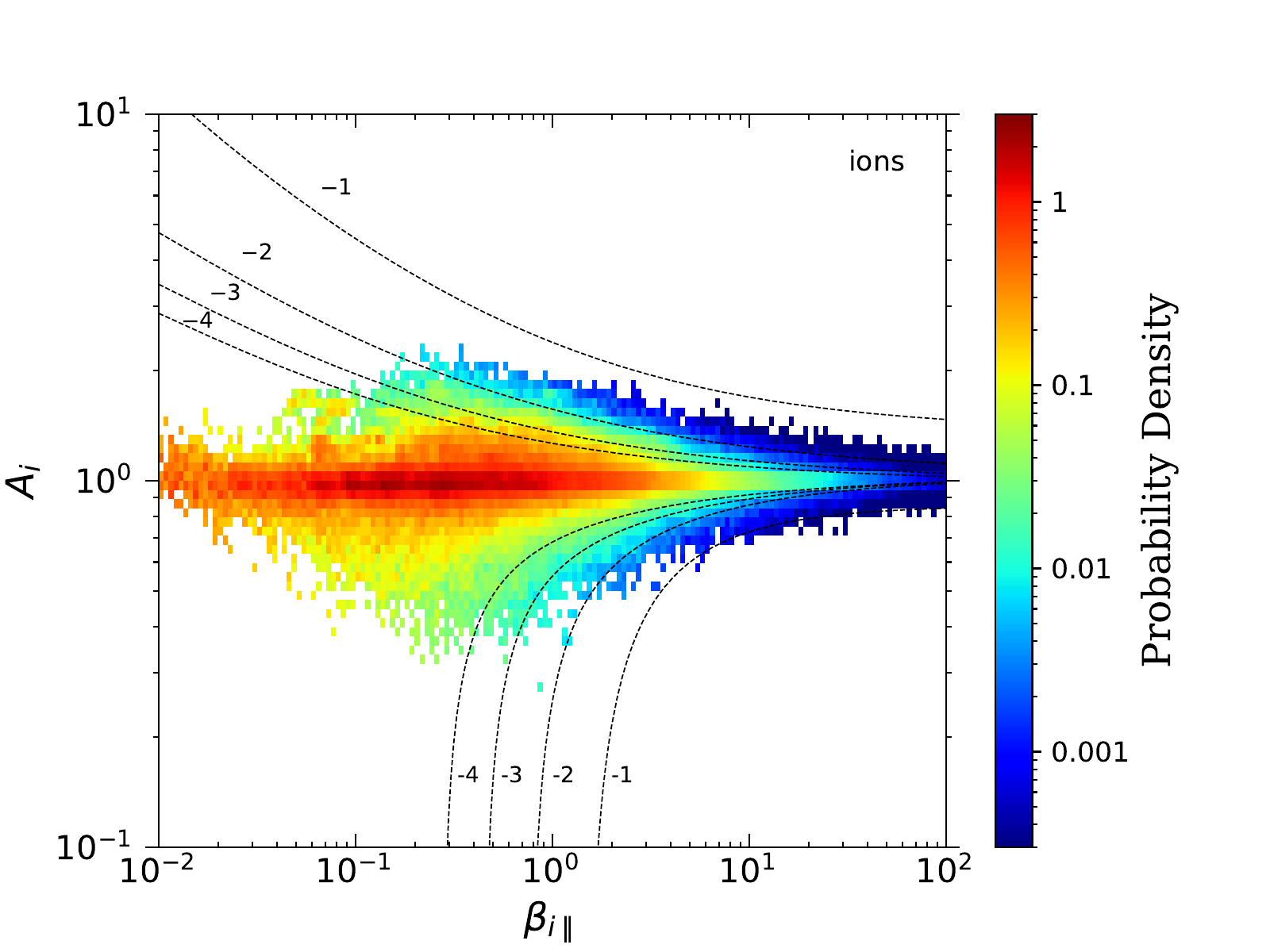} \hspace{0.1cm}
	\includegraphics[trim=0 0 105 10, clip, height=7.1cm]{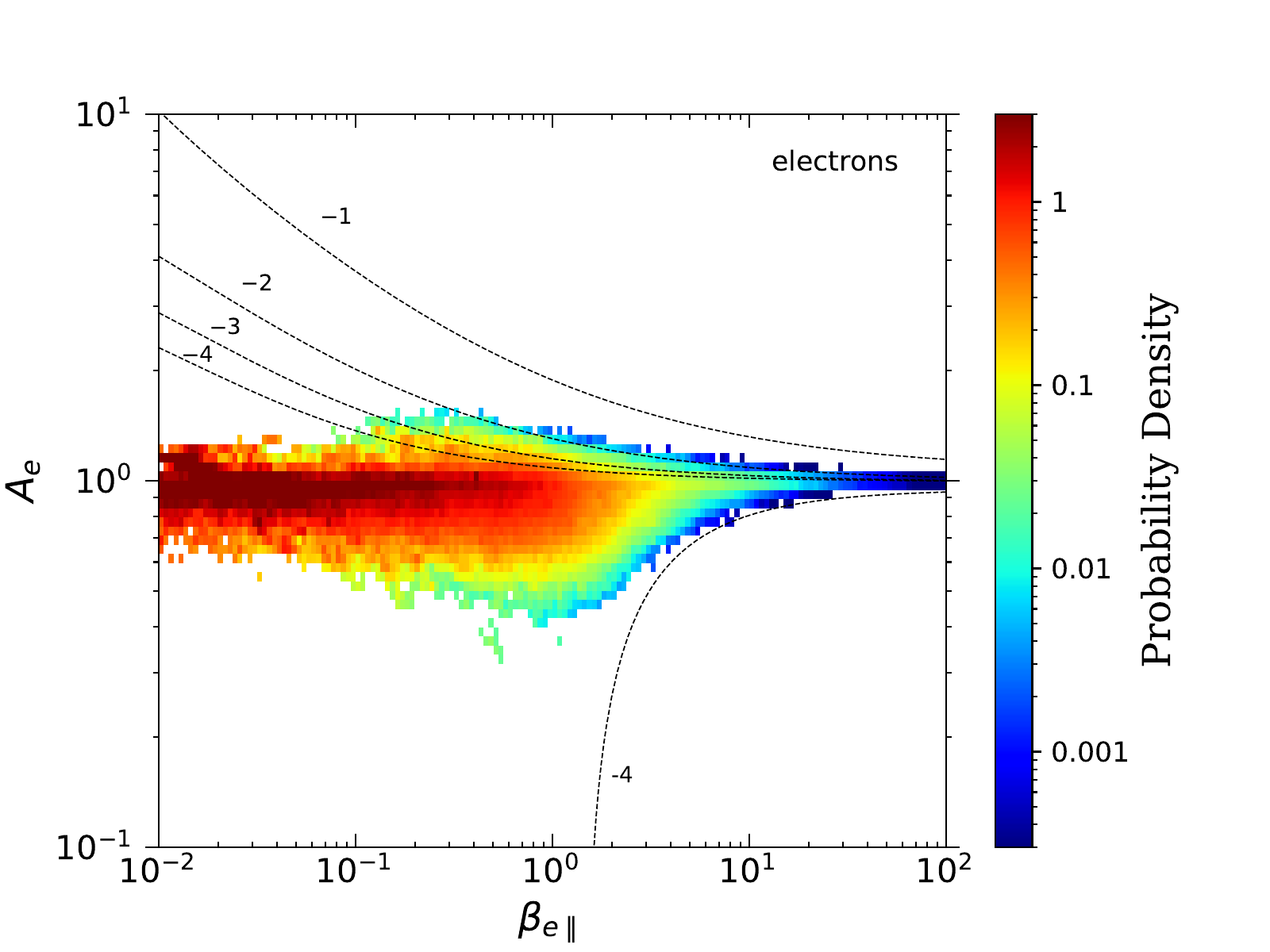} \hspace{0.4cm}
	\includegraphics[trim=360 0 16 10, clip, height=7.1cm]{density025e.pdf} \\
	\includegraphics[trim=0 0 105 10, clip, height=7.1cm]{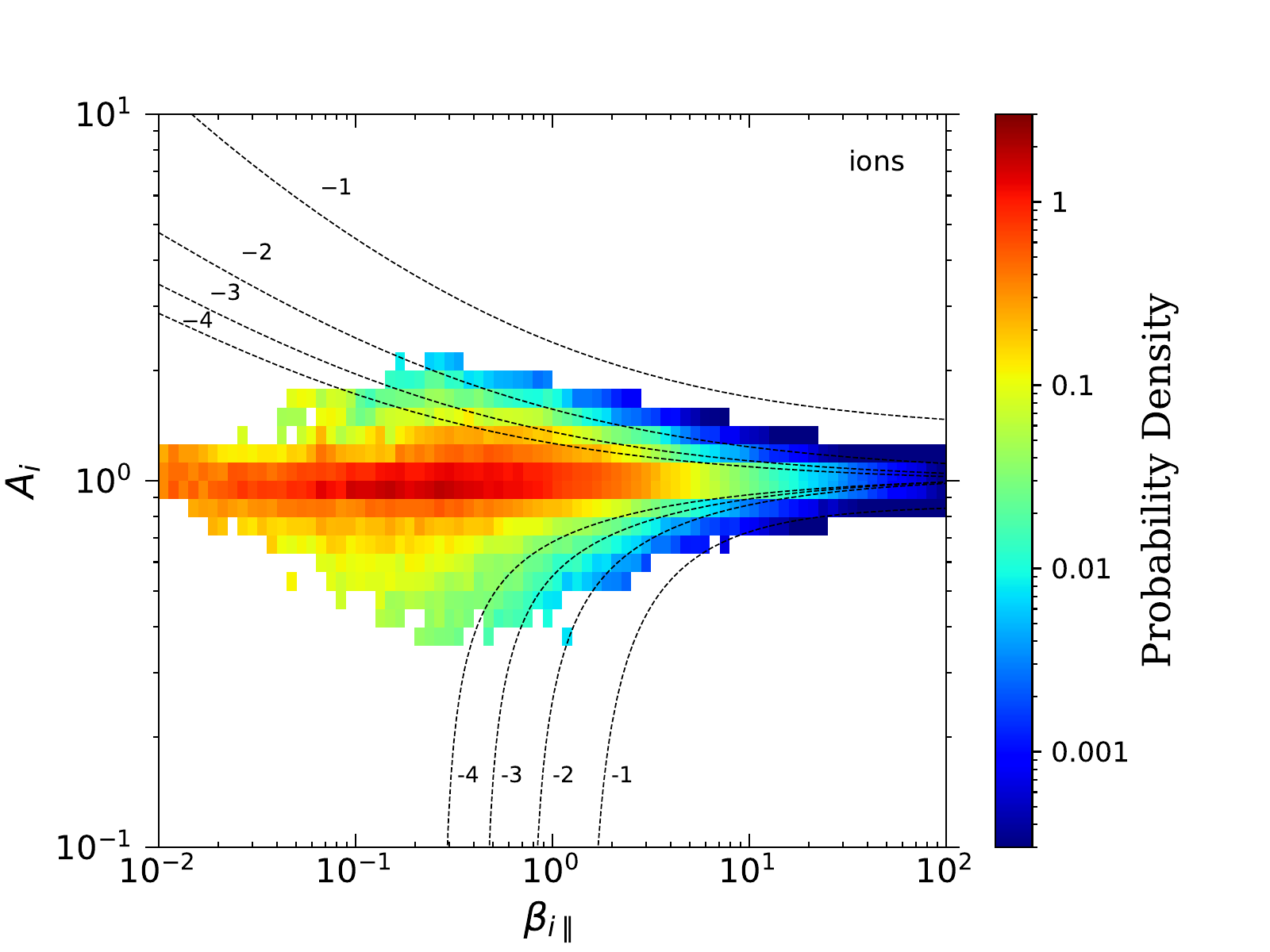} \hspace{0.1cm}
	\includegraphics[trim=0 0 105 10, clip, height=7.1cm]{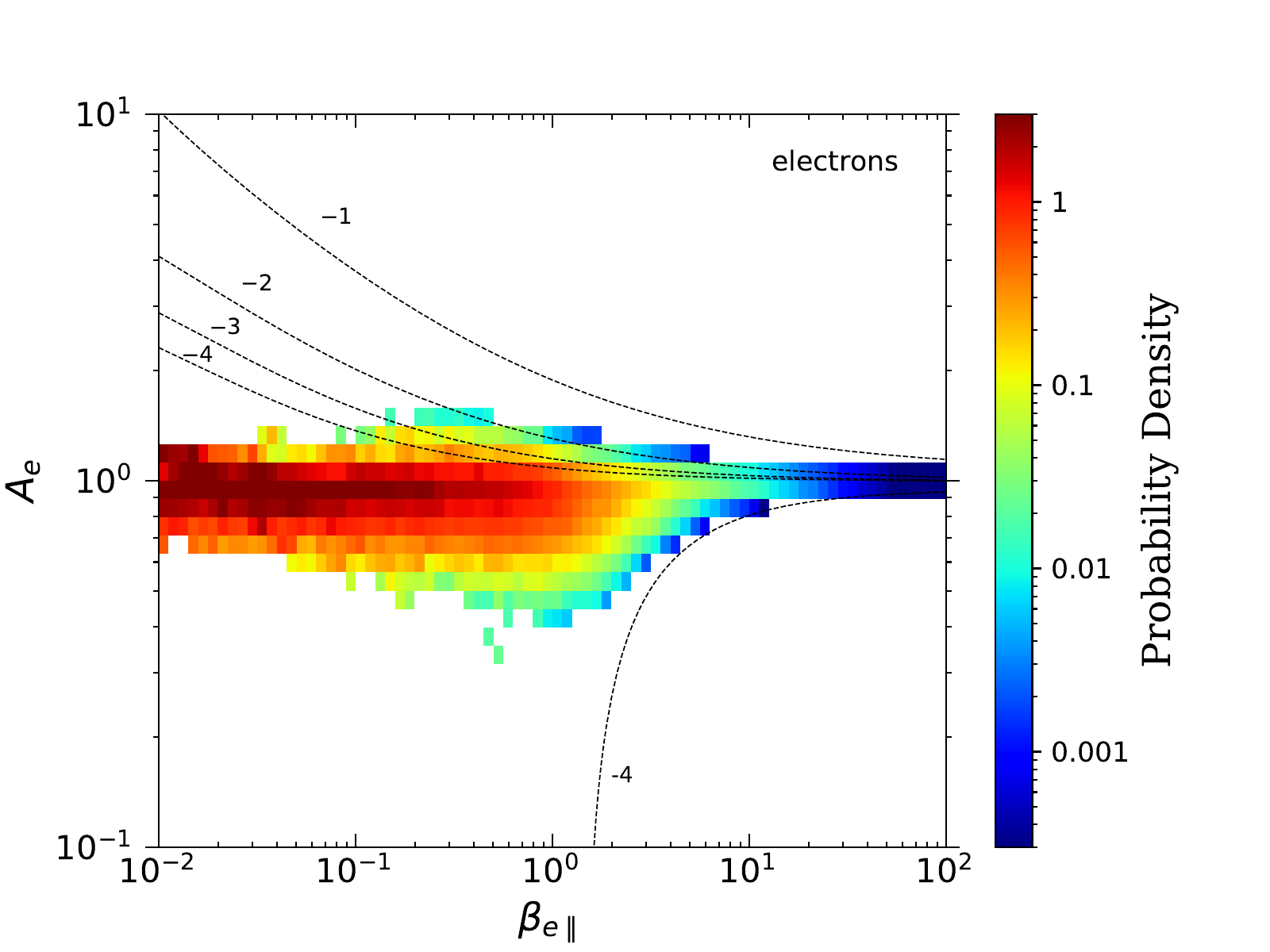} \hspace{0.4cm}
	\includegraphics[trim=360 0 16 10, clip, height=7.1cm]{density050e.pdf} \\
	\caption{Probability density of available magnetic fluctuation measurements in the plasma sheet between 2008 and 2009 for ions (left) and electrons (right), as a function of $\beta_\parallel$ and $A$.
	Top panels use log-log bins of $0.025\times0.025$ and cells are painted only if $N_M/N_T>10^{-5}$.
	Bottom panels use log-log bins of $0.05\times0.05$ and cells are painted only if $N_M/N_T>10^{-4.4}$.
	The segmented lines show the contours of the maximum growth rate of the instabilities listed in Tables \ref{tab:fiti} and \ref{tab:fite} (as in Fig. \ref{fig:bfluct}).
	The lines are labelled with the values of $\log_{10}(\gamma_{\rm{max}}/\Omega_i)$ (left plot) and $\log_{10}(\gamma_{\rm{max}}/\Omega_e)$ (right).
	\label{counts}}
\end{figure*}

In total, for the ions it was possible to calculate 576,954 magnetic fluctuations with their respective $A_i$ and $\beta_{i\parallel}$ values. For the electrons the total number of detections was similar: 569,450 magnetic fluctuations with $A_e$ and $\beta_{e\parallel}$ values. 
Fig.~\ref{counts} shows the probability density of the available measurements for each value of $\beta_\parallel$ and $A$, for both ions and electrons. 
This probability was calculated as the number of available measurements $N_M$ in a given bin divided by the total number of measurements $N_T$ for the particular species, and divided by the area of the bin (which is variable due to the logarithmic scale).
The top two panels in the figure were divided in cells of sizes $0.025\times0.025$ (in a log-log scale) and each box was colored only if $N_M/N_T>10^{-5}$. 
Similarly, the bottom two panels were divided in cells of sizes $0.05\times0.05$ and colored only if $N_M/N_T>10^{-4.4}$. 
Let us note that the change in threshold value between the two cases is consistent with an approximate increase by of factor of 4 in the number of measurements when going from a cell size of $0.025$ to $0.05$ in log-log space.
We further confirmed (not shown here) that these thresholds provide a consistent behavior when re-sampling the data into smaller sets with a third of the $N_T$ data points. 
Thus, we use a cell size of $0.025\times0.025$ with a threshold value of $N_M/N_T>10^{-5}$ in all other diagrams below.

The event probability patterns shown in Fig.~\ref{counts}(top) for each species are consistent with observations in the magnetosheath by the CLUSTER~\citep{Gary2005}, AMPTE~\citep{Phan1994,Anderson1994}, and MMS missions~\citep{mcg+18}; as well as in the solar wind~\citep{Gary2001ace,hpt+06,stm+08,avmw16,Huang2020}, computer simulations~\citep{Gary2001simul,Yoon2012}, and laboratory experiments~\citep{Scime2015,Scime2000,Beatty2020}

A key difference with most of these works is that the particle energy distributions for ions and electrons in the plasma sheet are better fitted by Kappa distributions \citep{esm+18,esem+21}, and that the ion temperature anisotropy is restricted to values $0.4\leq A_i\leq2$; whereas in the solar wind the ion velocity distributions are mostly Maxwellians with temperature anisotropies in the range $0.1\leq A_i\leq7$. 
On the other hand, the electron anisotropy distribution in the plasma sheet, as shown in Fig.~\ref{counts}, is similar to that observed for core and halo electrons in the solar wind \citep{stm+08}.

\begin{figure*}
	\includegraphics[trim=0 0 105 10, clip, height=7.0cm]{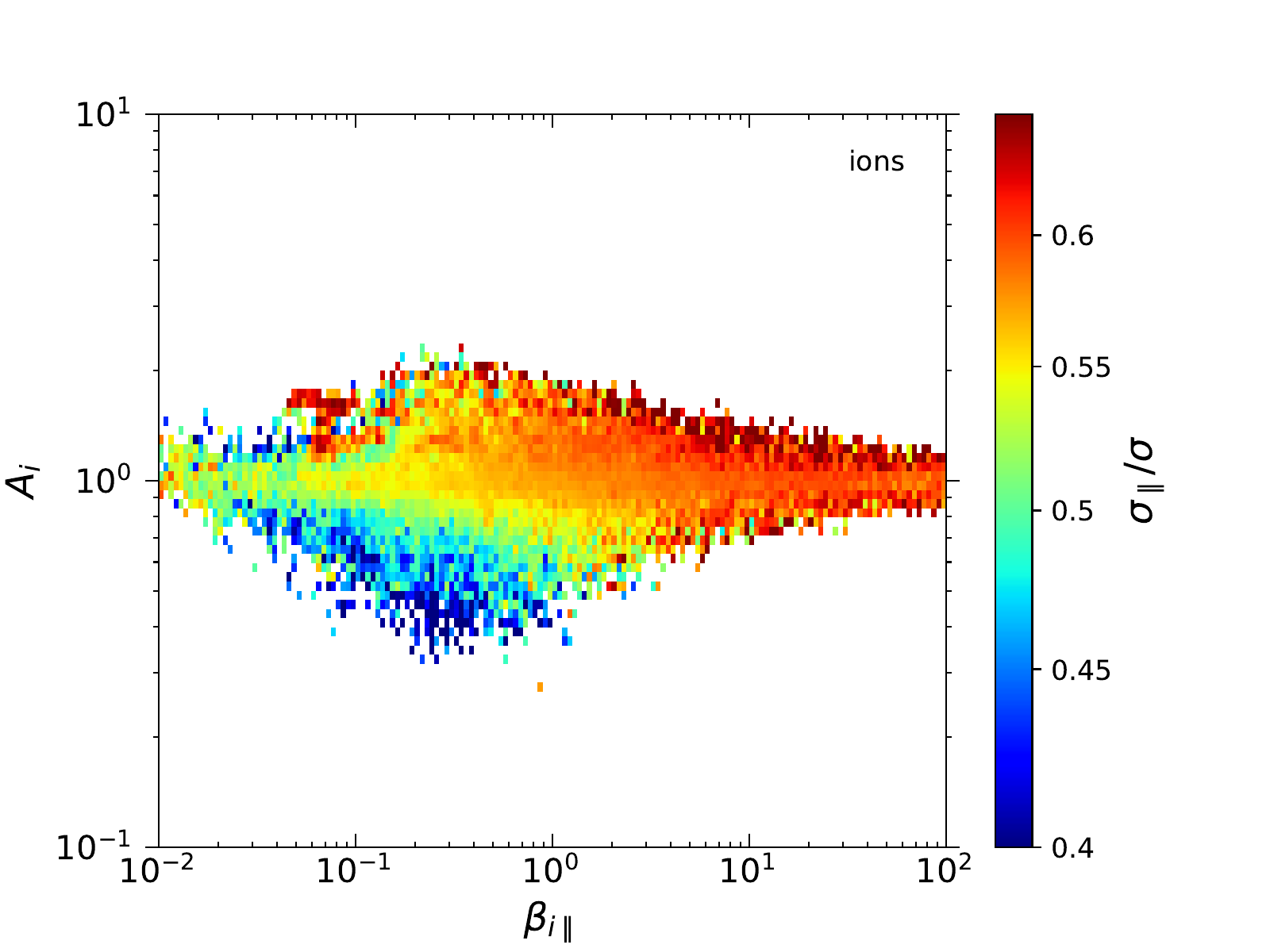} \hspace{0.1cm}
	\includegraphics[trim=0 0 105 10, clip, height=7.0cm]{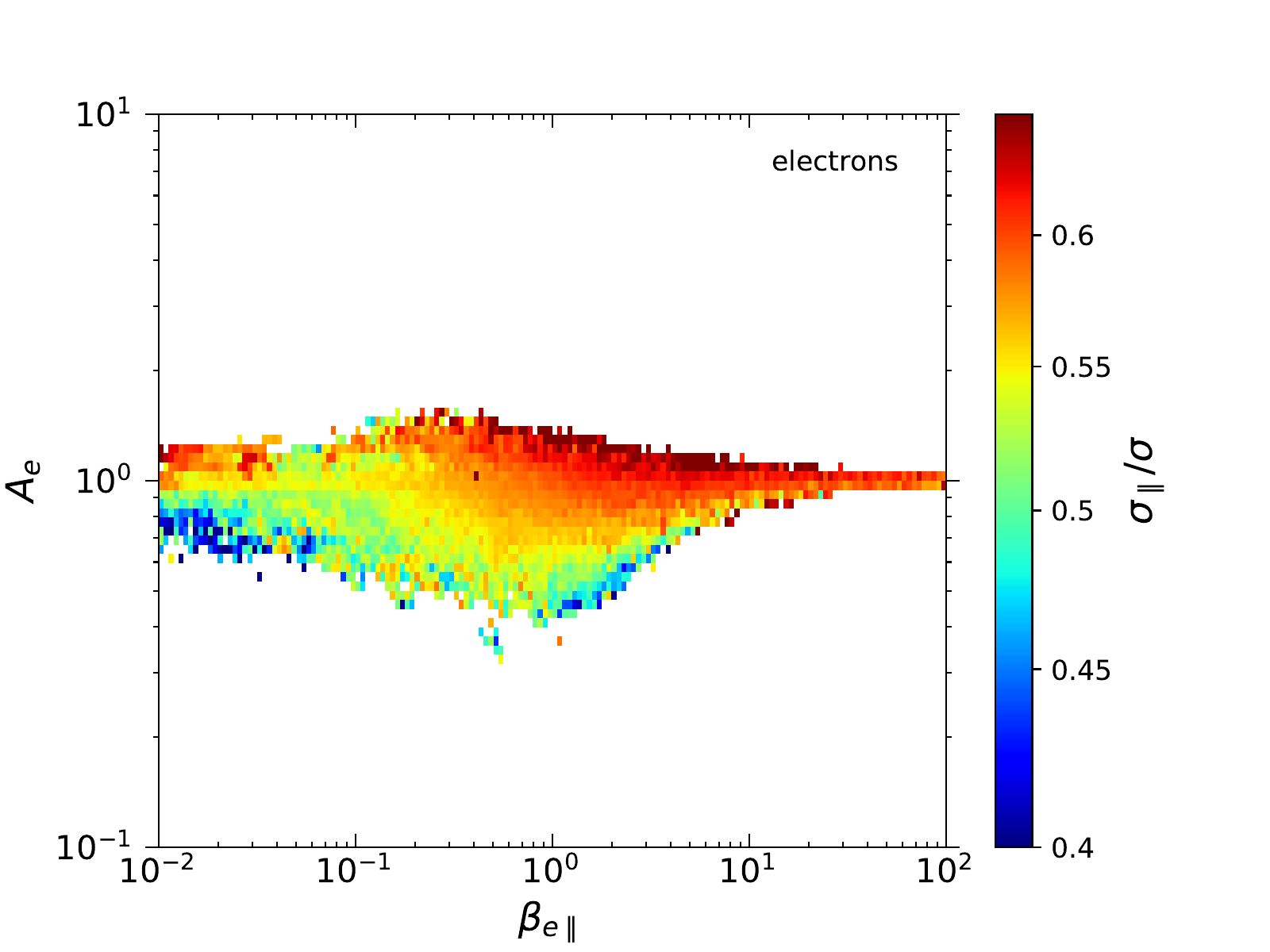} \hspace{0.4cm}
	\includegraphics[trim=360 0 16 10, clip, height=7.0cm]{SparaStote.pdf} \\
	\includegraphics[trim=0 0 105 10, clip, height=7.0cm]{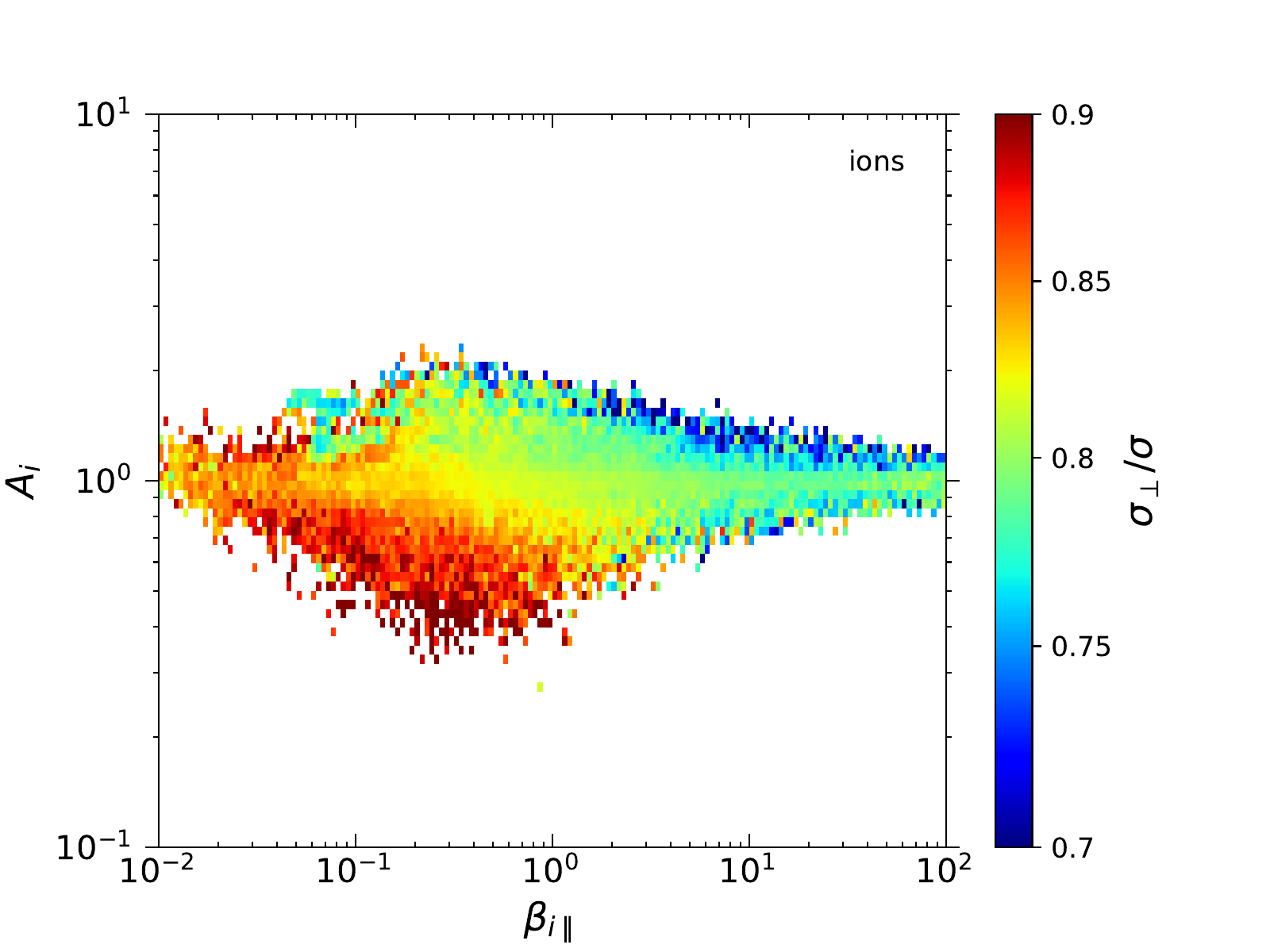} \hspace{0.1cm}
	\includegraphics[trim=0 0 105 10, clip, height=7.0cm]{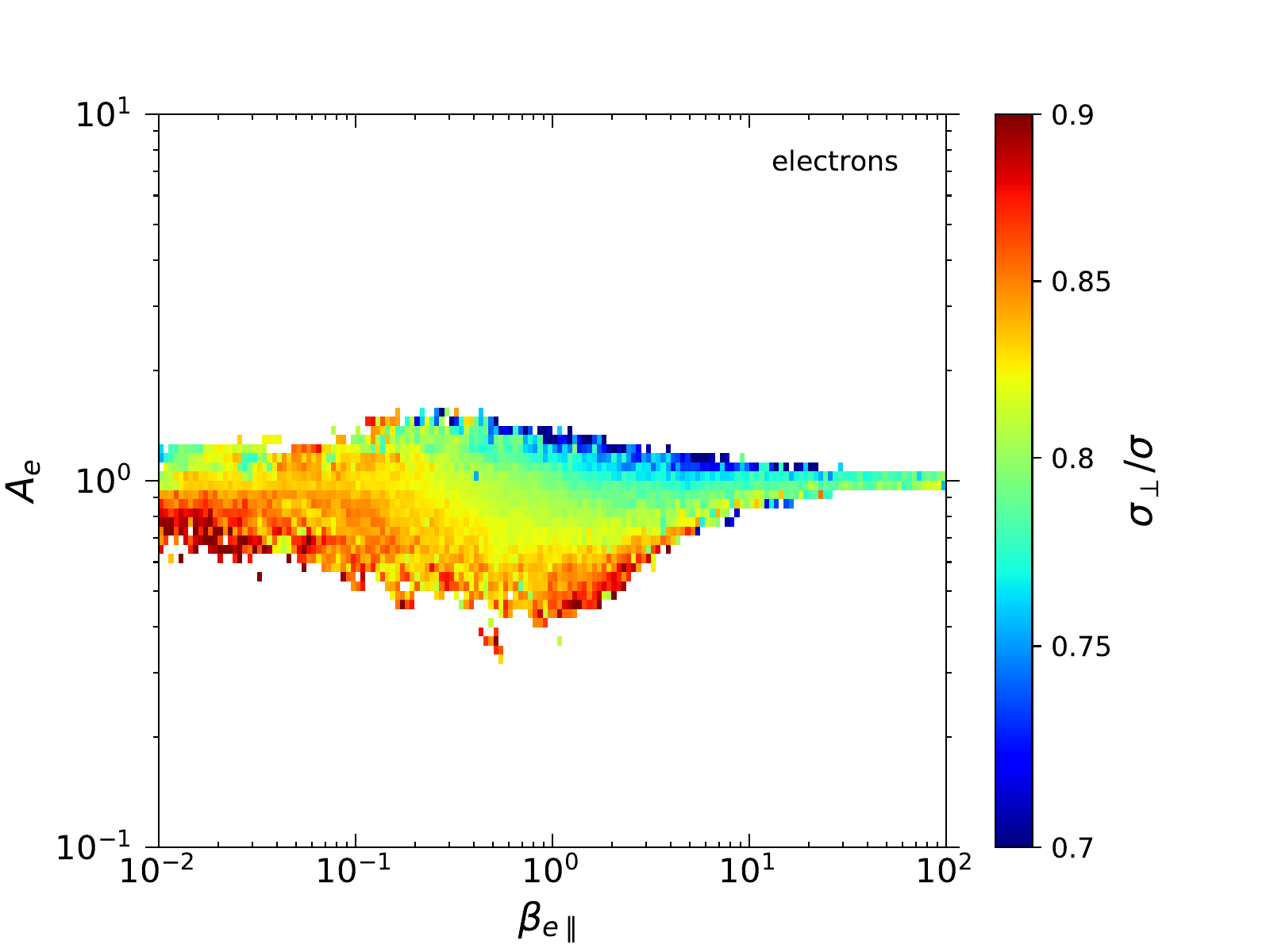} \hspace{0.4cm}
	\includegraphics[trim=360 0 16 10, clip, height=7.0cm]{SperStote.pdf} \\	
	\includegraphics[trim=0 0 105 10, clip, height=7.0cm]{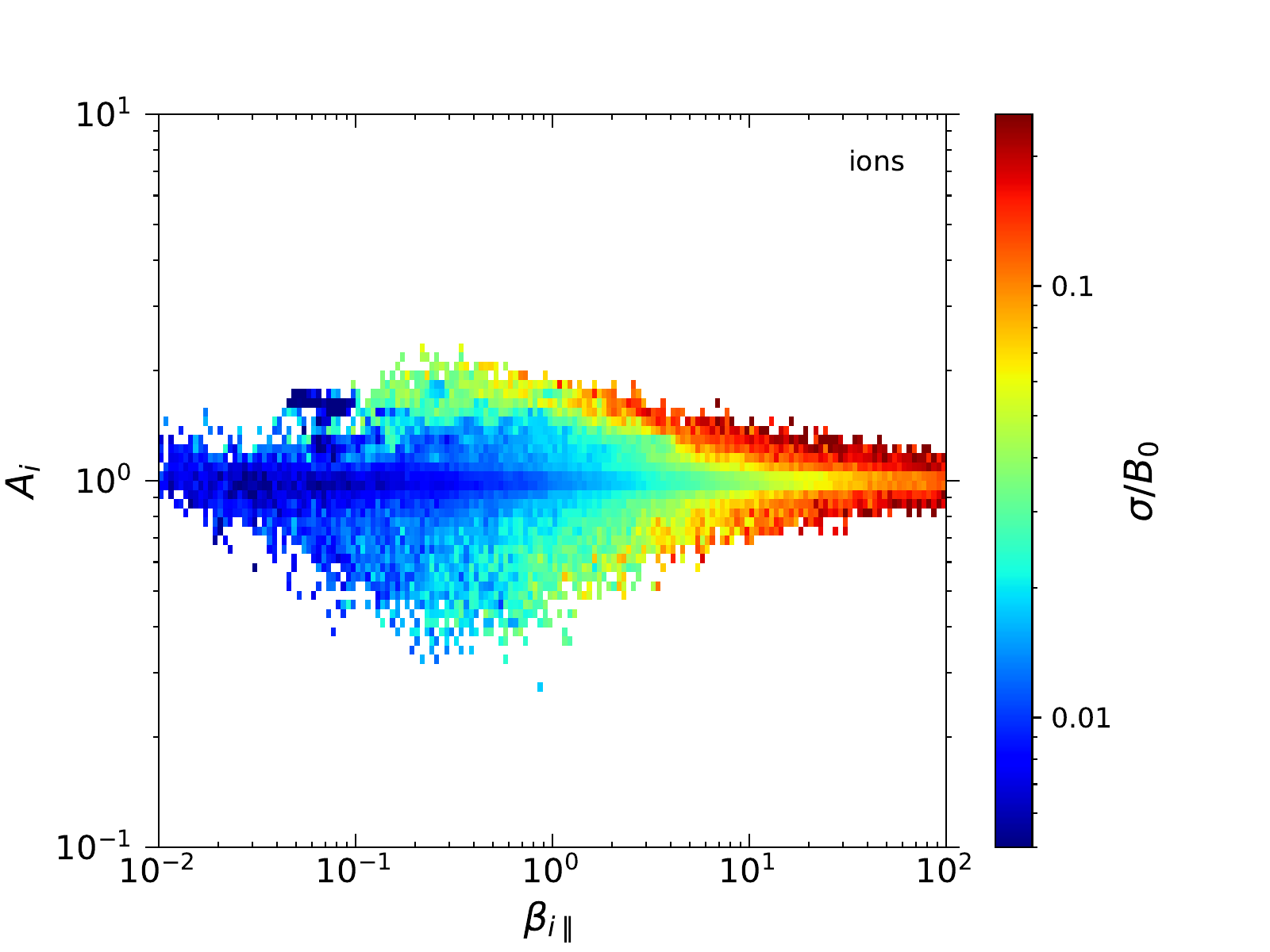} \hspace{0.1cm}
	\includegraphics[trim=0 0 105 10, clip, height=7.0cm]{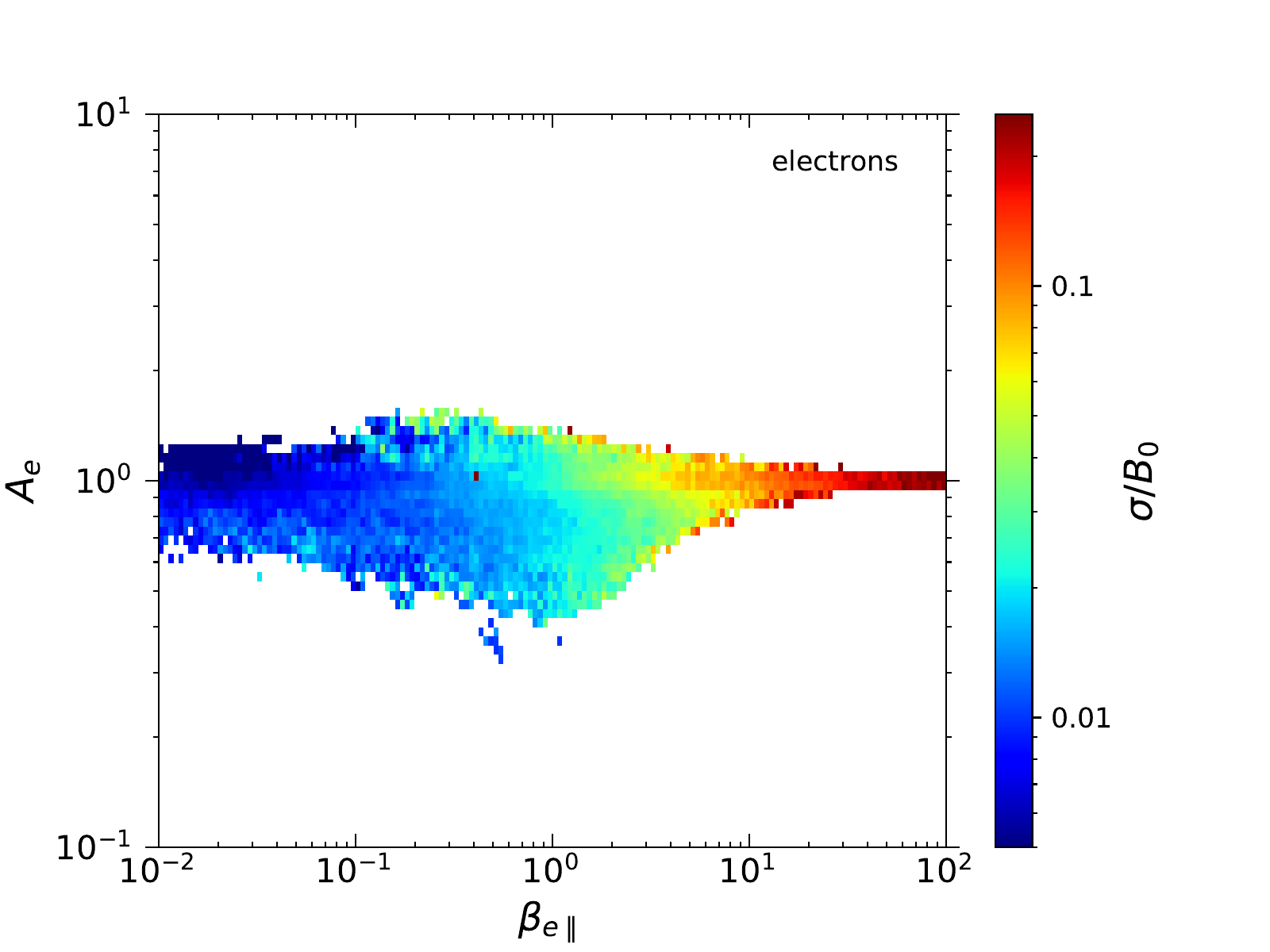} \hspace{0.4cm}
	\includegraphics[trim=360 0 16 10, clip, height=7.0cm]{stotB0e.pdf} \\
	\caption{$\beta_\parallel$-$A$ diagrams for ions (left) and electrons (right) showing the observed $\sigma_\parallel/\sigma$ (top), $\sigma_\perp/\sigma$ (middle), and $\sigma/B_0$ (bottom). 
	\label{f2}}
\end{figure*}

In Fig.~\ref{f2}, we display the observed parallel ($\sigma_\parallel/\sigma$) and perpendicular ($\sigma_\perp/\sigma$, known as magnetic compressibility) fluctuation levels, as defined by Eqs.~\eqref{eq:sigpal} and \eqref{eq:sigper}, respectively. 
We also show the total observed magnetic fluctuations ($\sigma/B_0$), for which high activity is observed even for $A\approx 1$, where one would (naively,  from a linear description) not expect significant fluctuations. 
We think that a relevant component of the observed fluctuations can be thermally induced, as suggested by \citet{nam+14}. 
To test such prediction for the magnetospheric environment we compare our THEMIS observations with the theory of spontaneous magnetic fluctuations outlined in Section~\ref{sec:theory}.

\begin{figure*}
	\plotone{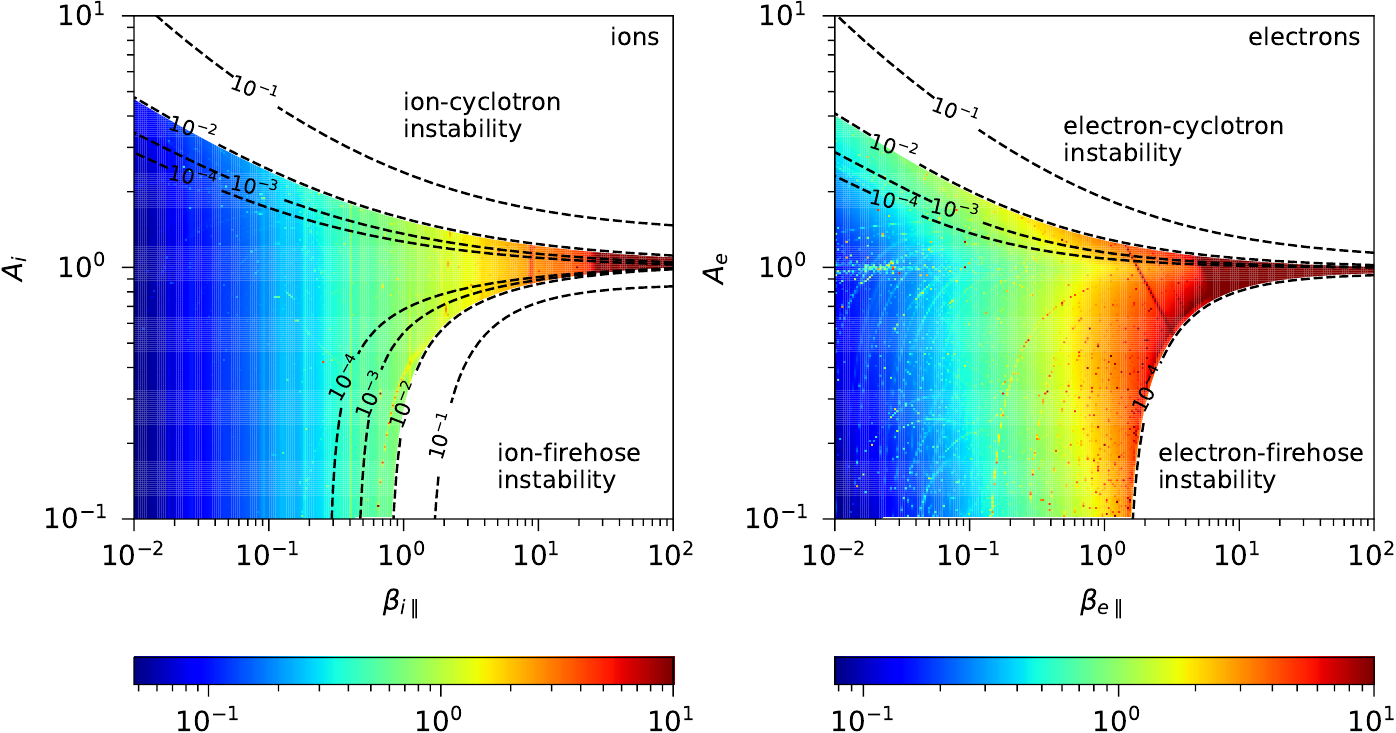}
	\caption{
	(colorbar) 
	Normalized fluctuating magnetic energy $\sqrt{W_B}$, Eq. \eqref{eq:WB}, in the $\beta_\parallel$-$A$ diagram of ions (left) and electrons (right). 
	We have used $\kappa_i=7$, $\kappa_e=5$,
    $\beta_{i\parallel}=5\beta_{e\parallel}$, and $A_e=1$ (left) and $A_i=1$ (right). 
    Dashed lines are the contours of the maximum growth-rate $\gamma_{\rm{}max}/\Omega_i$ of the ion-cyclotron ($A_i > 1$) and ion-firehose ($A_i < 1$) instabilities, in the case of the ions; 
    and $\gamma_{\rm{}max}/\Omega_e$ of the electron-cyclotron ($A_e>1$) and electron-firehose ($A_e<1$) instabilities, in the case of the electrons.
    In order to ensure resemblance with the observations, calculations of $W_B$ were suppressed (painted white) for parameters where
    $\gamma_{\rm{}max}/\Omega_i>10^{-2}$ in the diagram for the ions;
    and $\gamma_{\rm{}max}/\Omega_e>10^{-2}$ ($A_e>1$) and
    $\gamma_{\rm{}max}/\Omega_e>10^{-4}$ ($A_e<1$) for the electrons.
	\label{fig:bfluct}}
\end{figure*}

Figure~\ref{fig:bfluct} shows $W_B$ computed from Eq. \eqref{eq:WB} as a function of $\beta_{\parallel}$ and $A$ for ions and electrons.
It was assumed that $\kappa_e=5$, $\kappa_i=7$, and $\beta_{i\parallel}=5\beta_{e\parallel}$, as suggested by~\citet{esm+18}.
For fluctuations at the ion scales, we assumed that the electron anisotropy effects are negligible, for which we set $A_e=1$. 
Similarly, we set $A_i=1$ for fluctuations at the electron scales. 
The segmented lines in Figs.~\ref{fig:bfluct} and \ref{counts} are the contours of the maximum growth-rate of the instabilities listed in Tables \ref{tab:fiti} and \ref{tab:fite}. 
The pattern of $W_B$ in the $\beta_\parallel$-$A$ diagrams in Fig. \ref{fig:bfluct} can be compared with the observed  magnetic fluctuations $\sigma/B_0$ shown in Fig. \ref{f2} (bottom)  as a function of the proton and electron parameters.
The trending on the magnetic fluctuations increases mainly with $\beta_\parallel$, and to some extent with $A$.

\section{Discussion}
Although the total magnetic power in thermally induced electromagnetic fluctuations (calculated from Eq. \eqref{eq:WB} and shown in Fig. \ref{fig:bfluct}) is in arbitrary units, it resembles qualitatively well the patterns on the observed diagrams (obtained from THEMIS data and plotted in Fig. \ref{f2}), thereby suggesting that such fluctuations may constitute a relevant component of the turbulence observed.
The similarity of these results with those obtained for the solar wind \citep{nmm+2014} suggests that the kinetic physics plays an important role in the regulation of plasma turbulence and generation of fluctuations, in both the solar wind and the plasma sheet.

Figure~\ref{counts} shows that the majority of measurements lie far from the instability thresholds listed in Tables~\ref{tab:fiti} and \ref{tab:fite}. 
In particular, the ion-cyclotron (Alfv\'en) instability with values of $\gamma_{\rm{max}}/\Omega_i\approx0.01-0.1$ appears as an upper bound to the observable conditions for $A_i>1$; while the electron-anisotropy $A_e>1$ seems to be constrained by the electron-cyclotron (whistler) instability with $\gamma_{\rm max}/\Omega_e\approx0.01$ (or $\gamma_{\rm max}/\Omega_i\approx18.36$ if normalized to the ion gyrofrequency).
Thus, the anisotropy of both species for $A_\alpha>1$ seems to be constrained by approximately the same level (relative to the species scale) of the respective resonant instability.  
We acknowledge that this deserves further study.  
Of course, other instability rates $\gamma_{\rm max}$ would appear to represent the observations if a different threshold $N_M/N_T$ was used.
What matters is that the dependency on $\beta_\parallel$ of both observations and theory are similar, which suggests that the kinetic instabilities could indeed regulate the global behavior of the observations, hence of the turbulence.

Similarly, the electron temperature anisotropy $A_e<1$ seems to be constrained by a much lower level of the (non-resonant) electron-firehose instability compared with $A_e>1$, with $\gamma_{\rm max}/\Omega_e\approx10^{-4}$. For ions, it is not clear if a similar statement is true. While $A_i<1$ is indeed constrained by the (non-resonant) ion-firehose instability with $\gamma_{\rm max}/\Omega_i\approx10^{-1}$, this seems to apply only for high $\beta_{i\parallel}>10$ values. 
For $1<\beta_{i\parallel}<10$ other effects that were not considered here may affect the growth-rate levels. 
For instance, other values of the $\kappa$ indexes~\citep{nmm+15} and the presence of anisotropic electrons can affect the development of ion-firehose instabilities~\citep{Michno2014,Maneva2016}.

Let us note that the approximation of parallel propagation is quite reasonable, since $\sigma_\parallel/\sigma<1$ for most of the observed $A$ and $\beta_\parallel$ values, for both ions and electrons (Fig. \ref{f2}, top).
This is consistent with the propagation of non-compressive fluctuations (i.e. transverse ion- or electron-cyclotron waves propagating along the background magnetic field).

\section{Conclusions}
\label{sec:conclusions}
We have shown that temperature anisotropy at the kinetic level can regulate turbulence in the plasma sheet of the geomagnetic tail.
Similar results have been obtained for the solar wind~\citep{bkh+09} and the magnetosheath~\citep[e.g.][]{mcg+18}, and we conclude that this behavior might be more universal than previously expected. 
Furthermore, our results also suggest that physics at the kinetic level may be a relevant contributor to the magnetic fluctuations observed in these plasmas. We demonstrated that the resonant Alfv\'en and whistler instabilities may regulate the observed anisotropies for values above unity. 
Electron anisotropies $A_e<1$ seem to be constrained by the non-resonant firehose instability, while for ion anisotropies $A_i<1$ this statement is still not clear. 
The ion and electron cyclotron instabilities seem to operate on time scales comparable to the scales of protons or electrons. 
This does not seem to be the case for the firehose instabilities, so that more work and precise measurements are needed to understand these results.

In both the solar wind and the plasma sheet, the plasma appear to converge to a common dynamic quasi-equilibrium state. Understanding this universality may have implications for other astrophysical and laboratory plasma environments \citep{lnm+15}. 
Therefore, spontaneous fluctuations and their collisionless kinetic regulation may be fundamental features of space and astrophysical plasmas. 
Furthermore, this is a topic of current scientific interest, not only because of the need to advance our quantitative understanding of kinetic effects in the regulation of turbulence and production of electromagnetic fluctuations; but also because of its possible impact on the development of robust and accurate space weather forecasts during geomagnetic storms and substorms, where plasma turbulence and electromagnetic fluctuations seem to play a substantial role \citep[e.g.][]{bf03,kvv+00,szms00,vkv+03,sap+09,sa11,spva11,psav11,vtg+16,as+21}. 
Moreover, as \citet{nmm+2014}, we suggest that it may be possible to relate the thermal properties of the particle distribution functions with the production of magnetic fluctuations, a result that may position us one step closer to  understand  turbulent behavior in plasmas.

{\acknowledgments
This project has been financially supported by the AFOSR projects FA9550-19-1-0384 (M.S.; C.M.E.) and  FA9550-20-1-0189 (J.A.V.); by ANID/FONDECYT under the contracts 1191351 (P.S.M.), 1190703 (J.A.V.), 1211144 (M.S.), and 11180947 (R.E.N.); and by CONICYT-PAI 79170095 (R.E.N.). We also thank the support of CEDENNA under Grant AFB180001 (J.A.V.). We acknowledge NASA contract NAS5-02099 and V. Angelopoulos for the use of THEMIS mission data, specifically C.W. Carlson and J.P. McFadden for ESA data, D. Larson for SST data, and K.H. Glassmeier, U. Auster, and W. Baumjohann for FGM data. THEMIS satellite mission data used in this paper are available on the THEMIS mission website: \url{http://themis.ssl.berkeley.edu/index.shtml}.

}

%

\vspace*{5mm}
\facility{THEMIS}









\bibliography{plasma}

\begin{thebibliography}{}
\expandafter\ifx\csname natexlab\endcsname\relax\def\natexlab#1{#1}\fi
\providecommand{\url}[1]{\href{#1}{#1}}
\providecommand{\dodoi}[1]{doi:~\href{http://doi.org/#1}{\nolinkurl{#1}}}
\providecommand{\doeprint}[1]{\href{http://ascl.net/#1}{\nolinkurl{http://ascl.net/#1}}}
\providecommand{\doarXiv}[1]{\href{https://arxiv.org/abs/#1}{\nolinkurl{https://arxiv.org/abs/#1}}}

\bibitem[{Adrian {et~al.}(2016)Adrian, Vi{\~n}as, Moya, \& Wendel}]{avmw16}
Adrian, M.~L., Vi{\~n}as, A.~F., Moya, P.~S., \& Wendel, D.~E. 2016, The
  Astrophysical Journal, 833, 49, \dodoi{10.3847/1538-4357/833/1/49}

\bibitem[{Anderson {et~al.}(1994)Anderson, Fuselier, Gary, \&
  Denton}]{Anderson1994}
Anderson, B.~J., Fuselier, S.~A., Gary, S.~P., \& Denton, R.~E. 1994, J.
  Geophys. Res., 99, 5877, \dodoi{https://doi.org/10.1029/93JA02827}

\bibitem[{Angelopoulos(2008)}]{ang08}
Angelopoulos, V. 2008, \ssr, 141, 5, \dodoi{10.1007/s11214-008-9336-1}

\bibitem[{{Angelopoulos} {et~al.}(1993){Angelopoulos}, {Kennel}, {Coroniti},
  {Pellat}, {Spence}, {Kivelson}, {Walker}, {Baumjohann}, {Feldman}, {Gosling},
  \& {Russell}}]{akc+93}
{Angelopoulos}, V., {Kennel}, C.~F., {Coroniti}, F.~V., {et~al.} 1993,
  Geophysical Research Letters, 20, 1711, \dodoi{10.1029/93GL00847}

\bibitem[{{Antonova} \& {Ovchinnikov}(1999)}]{ao99}
{Antonova}, E.~E., \& {Ovchinnikov}, I.~L. 1999, Journal of Geophysical
  Research, 104, 17289, \dodoi{10.1029/1999JA900141}

\bibitem[{{Antonova} \& {Stepanova}(2021)}]{as+21}
{Antonova}, E.~E., \& {Stepanova}, M.~V. 2021, Frontiers in Astronomy and Space
  Sciences, 8, 26, \dodoi{10.3389/fspas.2021.622570}

\bibitem[{Araneda {et~al.}(2012)Araneda, Astudillo, \& Marsch}]{aam+12}
Araneda, J.~A., Astudillo, H., \& Marsch, E. 2012, Space Sci. Rev., 172, 361,
  \dodoi{10.1007/s11214-011-9773-0}

\bibitem[{{Auster} {et~al.}(2008){Auster}, {Glassmeier}, {Magnes}, {Aydogar},
  {Baumjohann}, {Constantinescu}, {Fischer}, {Fornacon}, {Georgescu}, {Harvey},
  {Hillenmaier}, {Kroth}, {Ludlam}, {Narita}, {Nakamura}, {Okrafka},
  {Plaschke}, {Richter}, {Schwarzl}, {Stoll}, {Valavanoglou}, \&
  {Wiedemann}}]{agm+08}
{Auster}, H.~U., {Glassmeier}, K.~H., {Magnes}, W., {et~al.} 2008, \ssr, 141,
  235, \dodoi{10.1007/s11214-008-9365-9}

\bibitem[{Bale {et~al.}(2009)Bale, Kasper, Howes, Quataert, Salem, \&
  Sundkvist}]{bkh+09}
Bale, S.~D., Kasper, J.~C., Howes, G.~G., {et~al.} 2009, Physical Review
  Letters, 103, 211101, \dodoi{10.1103/PhysRevLett.103.211101}

\bibitem[{Beatty {et~al.}(2020)Beatty, Steinberger, Aguirre, Beatty, Klein,
  McLaughlin, Neal, \& Scime}]{Beatty2020}
Beatty, C.~B., Steinberger, T.~E., Aguirre, E.~M., {et~al.} 2020, Phys.
  Plasmas, 27, 122101, \dodoi{10.1063/5.0029315}

\bibitem[{{Borovsky} {et~al.}(1997){Borovsky}, {Elphic}, {Funsten}, \&
  {Thomsen}}]{beft97}
{Borovsky}, J.~E., {Elphic}, R.~C., {Funsten}, H.~O., \& {Thomsen}, M.~F. 1997,
  Journal of Plasma Physics, 57, 1, \dodoi{10.1017/S0022377896005259}

\bibitem[{{Borovsky} \& {Funsten}(2003)}]{bf03}
{Borovsky}, J.~E., \& {Funsten}, H.~O. 2003, Journal of Geophysical Research
  (Space Physics), 108, 1284, \dodoi{10.1029/2002JA009625}

\bibitem[{Borovsky {et~al.}(2020)Borovsky, Delzanno, Valdivia, Moya, Stepanova,
  Birn, Blum, Lotko, \& Hesse}]{bdv+20}
Borovsky, J.~E., Delzanno, G.~L., Valdivia, J.~A., {et~al.} 2020, Journal of
  Atmospheric and Solar-Terrestrial Physics, 208, 105377,
  \dodoi{10.1016/j.jastp.2020.105377}

\bibitem[{{Denton} {et~al.}(2016){Denton}, {Borovsky}, {Stepanova}, \&
  {Valdivia}}]{dbsv16}
{Denton}, M.~H., {Borovsky}, J.~E., {Stepanova}, M., \& {Valdivia}, J.~A. 2016,
  Journal of Geophysical Research (Space Physics), 121, 10,783,
  \dodoi{10.1002/2016JA023362}

\bibitem[{Espinoza {et~al.}(2018)Espinoza, Stepanova, Moya, Antonova, \&
  Valdivia}]{esm+18}
Espinoza, C.~M., Stepanova, M., Moya, P.~S., Antonova, E.~E., \& Valdivia,
  J.~A. 2018, Geophysical Research Letters, 45, 6362,
  \dodoi{10.1029/2018GL078631}

\bibitem[{{Eyelade} {et~al.}(2021){Eyelade}, {Stepanova}, {Espinoza}, \&
  {Moya}}]{esem+21}
{Eyelade}, A.~V., {Stepanova}, M., {Espinoza}, C.~M., \& {Moya}, P.~S. 2021,
  Astrophys. J. Suppl. S., 253, 34, \dodoi{10.3847/1538-4365/abdec9}

\bibitem[{Gary {et~al.}(2005)Gary, Lavraud, Thomsen, Lefebvre, \&
  Schwartz}]{Gary2005}
Gary, S.~P., Lavraud, B., Thomsen, M.~F., Lefebvre, B., \& Schwartz, S.~J.
  2005, Geophys. Res. Lett., 32, \dodoi{https://doi.org/10.1029/2005GL023234}

\bibitem[{Gary {et~al.}(2001{\natexlab{a}})Gary, Skoug, Steinberg, \&
  Smith}]{Gary2001ace}
Gary, S.~P., Skoug, R.~M., Steinberg, J.~T., \& Smith, C.~W.
  2001{\natexlab{a}}, Geophys. Res. Lett., 28, 2759,
  \dodoi{https://doi.org/10.1029/2001GL013165}

\bibitem[{Gary {et~al.}(2001{\natexlab{b}})Gary, Yin, Winske, \&
  Ofman}]{Gary2001simul}
Gary, S.~P., Yin, L., Winske, D., \& Ofman, L. 2001{\natexlab{b}}, J. Geophys.
  Res., 106, 10715, \dodoi{https://doi.org/10.1029/2000JA000406}

\bibitem[{Hellberg \& Mace(2002)}]{hm02}
Hellberg, M.~A., \& Mace, R.~L. 2002, Physics of Plasmas, 9, 1495,
  \dodoi{10.1063/1.1462636}

\bibitem[{Hellinger {et~al.}(2006)Hellinger, Tr{\'a}vn{\'\i}{\v c}ek, Kasper,
  \& Lazarus}]{hpt+06}
Hellinger, P., Tr{\'a}vn{\'\i}{\v c}ek, P., Kasper, J.~C., \& Lazarus, A.~J.
  2006, Geophysical Research Letters, 33, L09101, \dodoi{10.1029/2006GL025925}

\bibitem[{Huang {et~al.}(2020)Huang, Kasper, Vech, Klein, Stevens,
  Martinovi{\'{c}}, Alterman, {\v{D}}urovcov{\'{a}}, Paulson, Maruca, Qudsi,
  Case, Korreck, Jian, Velli, Lavraud, Hegedus, Bert, Holmes, Bale, Larson,
  Livi, Whittlesey, Pulupa, MacDowall, Malaspina, Bonnell, Harvey, Goetz, \&
  de~Wit}]{Huang2020}
Huang, J., Kasper, J.~C., Vech, D., {et~al.} 2020, Astrophys. J. Suppl. Ser.,
  246, 70, \dodoi{10.3847/1538-4365/ab74e0}

\bibitem[{Isenberg {et~al.}(2013)Isenberg, Maruca, \& Kasper}]{imk13}
Isenberg, P.~A., Maruca, B.~A., \& Kasper, J.~C. 2013, ApJ, 773, 164,
  \dodoi{10.1088/0004-637X/773/2/164}

\bibitem[{Kasper {et~al.}(2006)Kasper, Lazarus, Steinberg, Ogilvie, \&
  Szabo}]{kls+06}
Kasper, J.~C., Lazarus, A.~J., Steinberg, J.~T., Ogilvie, K.~W., \& Szabo, A.
  2006, J. Geophys. Res., 111, A03105,
  \dodoi{https://doi.org/10.1029/2005JA011442}

\bibitem[{{Klimas} {et~al.}(2000){Klimas}, {Valdivia}, {Vassiliadis}, {Baker},
  {Hesse}, \& {Takalo}}]{kvv+00}
{Klimas}, A.~J., {Valdivia}, J.~A., {Vassiliadis}, D., {et~al.} 2000, Journal
  of Geophysical Research, 105, 18,765, \dodoi{10.1029/1999JA000319}

\bibitem[{{L{\'o}pez} {et~al.}(2015){L{\'o}pez}, {Navarro}, {Moya},
  {Vi{\~n}as}, {Araneda}, {Mu{\~n}oz}, \& {Alejandro Valdivia}}]{lnm+15}
{L{\'o}pez}, R.~A., {Navarro}, R.~E., {Moya}, P.~S., {et~al.} 2015, ApJ, 810,
  103, \dodoi{10.1088/0004-637X/810/2/103}

\bibitem[{Maneva {et~al.}(2016)Maneva, Lazar, Vi{\~{n}}as, \&
  Poedts}]{Maneva2016}
Maneva, Y., Lazar, M., Vi{\~{n}}as, A., \& Poedts, S. 2016, The Astrophysical
  Journal, 832, 64, \dodoi{10.3847/0004-637x/832/1/64}

\bibitem[{{Maruca} {et~al.}(2011){Maruca}, {Kasper}, \& {Bale}}]{mkb11}
{Maruca}, B.~A., {Kasper}, J.~C., \& {Bale}, S.~D. 2011, Physical Review
  Letters, 107, 201101, \dodoi{10.1103/PhysRevLett.107.201101}

\bibitem[{Maruca {et~al.}(2018)Maruca, Chasapis, Gary, Bandyopadhyay, Chhiber,
  Parashar, Matthaeus, Shay, Burch, Moore, Pollock, Giles, Paterson, Dorelli,
  Gershman, Torbert, Russell, \& Strangeway}]{mcg+18}
Maruca, B.~A., Chasapis, A., Gary, S.~P., {et~al.} 2018, The Astrophysical
  Journal, 866, 25, \dodoi{10.3847/1538-4357/aaddfb}

\bibitem[{McFadden {et~al.}(2008)McFadden, Carlson, Larson, Ludlam, Abiad,
  Elliott, Turin, Marckwordt, \& Angelopoulos}]{mcl+08}
McFadden, J., Carlson, C., Larson, D., {et~al.} 2008, Space Science Reviews,
  141, 277, \dodoi{10.1007/s11214-008-9440-2}

\bibitem[{Michno {et~al.}(2014)Michno, Lazar, Yoon, \&
  Schlickeiser}]{Michno2014}
Michno, M.~J., Lazar, M., Yoon, P.~H., \& Schlickeiser, R. 2014, The
  Astrophysical Journal, 781, 49, \dodoi{10.1088/0004-637x/781/1/49}

\bibitem[{Moya {et~al.}(2020)Moya, Lazar, \& Poedts}]{m+2020}
Moya, P.~S., Lazar, M., \& Poedts, S. 2020, Plasma Physics and Controlled
  Fusion, 63, 025011, \dodoi{10.1088/1361-6587/abce1a}

\bibitem[{Moya \& Navarro(2021)}]{mn2021}
Moya, P.~S., \& Navarro, R.~E. 2021, Frontiers in Physics, 9, 175,
  \dodoi{10.3389/fphy.2021.624748}

\bibitem[{Navarro {et~al.}(2014{\natexlab{a}})Navarro, Araneda, {Mu{\~n}oz},
  {Moya}, {Vi{\~n}as}, \& {Valdivia}}]{nam+14}
Navarro, R., Araneda, J., {Mu{\~n}oz}, V., {et~al.} 2014{\natexlab{a}}, Physics
  of Plasmas, 21, 092902, \dodoi{10.1063/1.4894700}

\bibitem[{Navarro {et~al.}(2014{\natexlab{b}})Navarro, Moya, Mu{\~n}oz,
  Araneda, F.-Vi{\~n}as, \& Valdivia}]{nmm+2014}
Navarro, R., Moya, P., Mu{\~n}oz, V., {et~al.} 2014{\natexlab{b}}, Physical
  Review Letters, 112, 245001, \dodoi{10.1103/PhysRevLett.112.245001}

\bibitem[{Navarro {et~al.}(2015)Navarro, Moya, Mu{\~n}oz, Araneda,
  F.-Vi{\~n}as, \& Valdivia}]{nmm+15}
Navarro, R.~E., Moya, P.~S., Mu{\~n}oz, V., {et~al.} 2015, Journal of
  Geophysical Research (Space Physics), 120, 2382, \dodoi{10.1002/2014JA020550}

\bibitem[{Ovchinnikov {et~al.}(2000)Ovchinnikov, Antonova, \&
  Yermolaev}]{oay00}
Ovchinnikov, I.~L., Antonova, E.~E., \& Yermolaev, Y.~I. 2000, Cosmic Research,
  38, 557, \dodoi{10.1023/A:1026686600686}

\bibitem[{Phan {et~al.}(1994)Phan, Paschmann, Baumjohann, Sckopke, \&
  Lühr}]{Phan1994}
Phan, T.~D., Paschmann, G., Baumjohann, W., Sckopke, N., \& Lühr, H. 1994, J.
  Geophys. Res., 99, 121, \dodoi{https://doi.org/10.1029/93JA02444}

\bibitem[{{Pinto} {et~al.}(2011){Pinto}, {Stepanova}, {Antonova}, \&
  {Valdivia}}]{psav11}
{Pinto}, V., {Stepanova}, M., {Antonova}, E.~E., \& {Valdivia}, J.~A. 2011,
  Journal of Atmospheric and Solar-Terrestrial Physics, 73, 1472,
  \dodoi{10.1016/j.jastp.2011.05.007}

\bibitem[{Scime {et~al.}(2000)Scime, Keiter, Balkey, Boivin, Kline, Blackburn,
  \& Gary}]{Scime2000}
Scime, E.~E., Keiter, P.~A., Balkey, M.~M., {et~al.} 2000, Phys. Plasmas, 7,
  2157, \dodoi{10.1063/1.874036}

\bibitem[{Scime {et~al.}(2015)Scime, Keiter, Balkey, Kline, Sun, Keesee,
  Hardin, Biloiu, Houshmandyar, Chakraborty~Thakur, \& et~al.}]{Scime2015}
---. 2015, J. Plasma Phys., 81, 345810103, \dodoi{10.1017/S0022377814000890}

\bibitem[{{Sitnov} {et~al.}(2000){Sitnov}, {Zelenyi}, {Malova}, \&
  {Sharma}}]{szms00}
{Sitnov}, M.~I., {Zelenyi}, L.~M., {Malova}, H.~V., \& {Sharma}, A.~S. 2000,
  Journal of Geophysical Research, 105, 13029, \dodoi{10.1029/1999JA000431}

\bibitem[{{Stepanova} \& {Antonova}(2011)}]{sa11}
{Stepanova}, M., \& {Antonova}, E.~E. 2011, Journal of Atmospheric and
  Solar-Terrestrial Physics, 73, 1636, \dodoi{10.1016/j.jastp.2011.02.009}

\bibitem[{{Stepanova} {et~al.}(2009){Stepanova}, {Antonova}, {Paredes-Davis},
  {Ovchinnikov}, \& {Yermolaev}}]{sap+09}
{Stepanova}, M., {Antonova}, E.~E., {Paredes-Davis}, D., {Ovchinnikov}, I.~L.,
  \& {Yermolaev}, Y.~I. 2009, Annales Geophysicae, 27, 1407,
  \dodoi{10.5194/angeo-27-1407-2009}

\bibitem[{{Stepanova} {et~al.}(2011){Stepanova}, {Pinto}, {Valdivia}, \&
  {Antonova}}]{spva11}
{Stepanova}, M., {Pinto}, V., {Valdivia}, J.~A., \& {Antonova}, E.~E. 2011,
  Journal of Geophysical Research (Space Physics), 116, 0,
  \dodoi{10.1029/2010JA015887}

\bibitem[{Summers \& Thorne(1991)}]{st91}
Summers, D., \& Thorne, R.~M. 1991, Physics of Fluids B: Plasma Physics, 3,
  1835, \dodoi{10.1063/1.859653}

\bibitem[{{Valdivia} {et~al.}(2003){Valdivia}, {Klimas}, {Vassiliadis},
  {Uritsky}, \& {Takalo}}]{vkv+03}
{Valdivia}, J.~A., {Klimas}, A., {Vassiliadis}, D., {Uritsky}, V., \& {Takalo},
  J. 2003, Space Science Reviews, 107, 515, \dodoi{10.1023/A:1025518527128}

\bibitem[{Valdivia {et~al.}(2016)Valdivia, Toledo, Gallo, Mu{\~n}oz, Rogan,
  Stepanova, Moya, Navarro, Vi{\~n}as, Araneda, L{\'o}pez, \&
  D{\'\i}az}]{vtg+16}
Valdivia, J.~A., Toledo, B.~A., Gallo, N., {et~al.} 2016, Advances in Space
  Research, 58, 2126, \dodoi{10.1016/j.asr.2016.04.017}

\bibitem[{{Vi{\~n}as} {et~al.}(2015){Vi{\~n}as}, {Moya}, {Navarro}, {Valdivia},
  {Araneda}, \& {Mu{\~n}oz}}]{vmn+15}
{Vi{\~n}as}, A.~F., {Moya}, P.~S., {Navarro}, R.~E., {et~al.} 2015, Journal of
  Geophysical Research (Space Physics), 120, 3307, \dodoi{10.1002/2014JA020554}

\bibitem[{Vi{\~n}as {et~al.}(2017)Vi{\~n}as, Gaelzer, Moya, Mace, \&
  Araneda}]{Vinas2017}
Vi{\~n}as, A.~F., Gaelzer, R., Moya, P.~S., Mace, R., \& Araneda, J.~A. 2017,
  in Kappa {Distributions}, ed. G.~Livadiotis (Elsevier), 329--361,
  \dodoi{10.1016/B978-0-12-804638-8.00007-3}

\bibitem[{{Volwerk} {et~al.}(2004){Volwerk}, {V{\"o}r{\"o}s}, {Baumjohann},
  {Nakamura}, {Runov}, {Zhang}, {Glassmeier}, {Treumann}, {Klecker}, {Balogh},
  \& {R{\`e}me}}]{vvb+04}
{Volwerk}, M., {V{\"o}r{\"o}s}, Z., {Baumjohann}, W., {et~al.} 2004, Annales
  Geophysicae, 22, 2525, \dodoi{10.5194/angeo-22-2525-2004}

\bibitem[{{V{\"o}r{\"o}s} {et~al.}(2004){V{\"o}r{\"o}s}, {Baumjohann},
  {Nakamura}, {Volwerk}, {Runov}, {Zhang}, {Eichelberger}, {Treumann},
  {Georgescu}, {Balogh}, {Klecker}, \& {R{\'e}Me}}]{vbn+04}
{V{\"o}r{\"o}s}, Z., {Baumjohann}, W., {Nakamura}, R., {et~al.} 2004, Journal
  of Geophysical Research (Space Physics), 109, A11215,
  \dodoi{10.1029/2004JA010404}

\bibitem[{\v{S}tver\'{a}k {et~al.}(2008)\v{S}tver\'{a}k, Tr\'{a}vn\'{i}\v{c}ek,
  Maksimovic, Marsch, Fazakerley, \& Scime}]{stm+08}
\v{S}tver\'{a}k, v., Tr\'{a}vn\'{i}\v{c}ek, P., Maksimovic, M., {et~al.} 2008,
  J. Geophys. Res., 113, A03103, \dodoi{https://doi.org/10.1029/2007JA012733}

\bibitem[{{Weygand} {et~al.}(2005){Weygand}, {Kivelson}, {Khurana}, {Schwarzl},
  {Thompson}, {McPherron}, {Balogh}, {Kistler}, {Goldstein}, {Borovsky}, \&
  {Roberts}}]{wkk+05}
{Weygand}, J.~M., {Kivelson}, M.~G., {Khurana}, K.~K., {et~al.} 2005, Journal
  of Geophysical Research (Space Physics), 110, A01205,
  \dodoi{10.1029/2004JA010581}

\bibitem[{Yoon \& Seough(2012)}]{Yoon2012}
Yoon, P.~H., \& Seough, J. 2012, J. Geophys. Res., 117,
  \dodoi{https://doi.org/10.1029/2012JA017697}

\bibitem[{Yue {et~al.}(2016)Yue, An, Bortnik, Ma, Li, Thorne, Reeves,
  Gkioulidou, Mitchell, \& Kletzing}]{yab+16}
Yue, C., An, X., Bortnik, J., {et~al.} 2016, Geophysical Research Letters, 43,
  7804, \dodoi{https://doi.org/10.1002/2016GL070084}

\end{thebibliography}



\end{document}